\documentclass[11pt]{article}
\usepackage{cite}
\usepackage{mathrsfs}
\usepackage{amssymb,amsfonts,amsmath}
\usepackage{enumerate}
\usepackage{indentfirst}
\usepackage{latexsym,bm}
\usepackage[noend]{algpseudocode}
\usepackage{algorithmicx,algorithm}
\usepackage{algorithm}
\usepackage{makeidx}
\usepackage{fancybox}
\usepackage{color}
\usepackage{multicol}
\usepackage[latin1]{inputenc}
\usepackage{graphicx}
\usepackage{pstricks,pst-node,pst-text,pst-3d}
\usepackage{tikz}
\newtheorem{thm}{Theorem}[section]

\newtheorem{lem}[thm]{Lemma}

\newtheorem{conj}{Conjecture}[section]

\newenvironment{pf}{{\noindent \it \bf Proof:}}{{\hfill$\Box$}\\}

\begin{document}

\title{\bf Arc-disjoint strong spanning subdigraphs in compositions and products  of digraphs}
\author{Yuefang Sun$^{1,}$\footnote{Yuefang Sun was supported by National Natural Science Foundation of China (No. 11401389).}{ } Gregory Gutin$^{2,}$\footnote{Corresponding author. Gregory Gutin was partially supported by Royal Society Wolfson Research Merit Award.} 
{ } Jiangdong Ai$^{2}$\\
$^{1}$ Department of Mathematics,
Shaoxing University\\
Zhejiang 312000, P. R. China, yuefangsun2013@163.com\\
$^{2}$ Department of Computer Science\\
Royal Holloway, University of London\\
Egham, Surrey, TW20 0EX, UK\\
g.gutin@rhul.ac.uk, ai18810910153@163.com}
\date{}
\maketitle

\begin{abstract}
A digraph $D=(V,A)$ has a good decomposition if $A$ has two disjoint sets $A_1$ and $A_2$ such that both
$(V,A_1)$ and $(V,A_2)$ are strong.
Let $T$ be a digraph with $t$ vertices $u_1,\dots , u_t$ and let $H_1,\dots H_t$ be digraphs such that $H_i$ has vertices $u_{i,j_i},\ 1\le j_i\le n_i.$
Then the composition $Q=T[H_1,\dots , H_t]$ is a digraph with vertex set $\{u_{i,j_i}\mid 1\le i\le t, 1\le j_i\le n_i\}$ and arc set
$$A(Q)=\cup^t_{i=1}A(H_i)\cup \{u_{ij_i}u_{pq_p}\mid u_iu_p\in A(T), 1\le j_i\le n_i, 1\le q_p\le n_p\}.$$ 

For digraph compositions $Q=T[H_1,\dots H_t]$, we obtain sufficient conditions for 
$Q$ to have a good decomposition and a characterization
of $Q$ with a good decomposition when $T$ is a strong semicomplete digraph and 
each $H_i$ is an arbitrary digraph with at least two vertices.

For digraph products, we prove the following: (a) if $k\geq 2$ is an integer and $G$ is a strong digraph which has a collection of arc-disjoint cycles covering all 
vertices, then the Cartesian product digraph $G^{\square k}$ (the $k$th powers with respect to Cartesian product) has a good decomposition;
(b) for any strong digraphs $G, H$, the strong product $G\boxtimes H$ has a good decomposition. 
\vspace{0.3cm}\\
{\bf Keywords:} strong spanning subdigraph; decomposition into
strong spanning subdigraphs; semicomplete digraph; digraph composition; Cartesian product; strong product.
\vspace{0.3cm}\\ {\bf AMS subject
classification (2010)}: 05C20, 05C70, 05C76, 05C85.

\end{abstract}

\section{Introduction}
We refer the readers to \cite{Bang-Jensen-Gutin, Bang-Jensen-Gutin2,
Bondy} for graph theoretical notation and terminology not given
here. A digraph $D = (V, A)$ is {\em strongly connected} (or {\em
strong}) if there exists a path from $x$ to $y$ and a path from $y$
to $x$ in $D$ for every pair of distinct vertices $x, y$ of $D$. A
digraph $D$ is {\em $k$-arc-strong} if $D- X$ is strong for every
subset $X\subseteq A$ of size at most $k- 1$. 

An {\em out-branching} $B^+_s$ (respectively, {\em in-branching}
$B^-_s$) in a digraph $D = (V, A)$ is a connected spanning
subdigraph of $D$ in which each vertex $x\neq s$ has precisely one
arc entering (leaving) it and $s$ has no arcs entering (leaving) it.
The vertex $s$ is the root of $B^+_s$ (respectively, $B^-_s$). 
Edmonds \cite{Edmonds} characterized digraphs with have $k$
arc-disjoint out-branchings rooted at a specified vertex $s.$
Furthermore, there exists a polynomial algorithm for finding $k$
arc-disjoint out-branchings from a given root $s$ if they exist (see
p. 346 of \cite{Bang-Jensen-Gutin}). However, if we ask for the
existence of a pair of arc-disjoint branchings $B^+_s$, $B^-_s$ such
that the first is an out-branching rooted at $s$ and the latter is
an in-branching rooted at $s$, then the problem becomes NP-complete
(see Section 9.6 of \cite{Bang-Jensen-Gutin}). In connection with
this problem, Thomassen \cite{Thomassen} posed the following conjecture:
There exists an integer $N$ so that every $N$-arc-strong digraph $D$
contains a pair of arc-disjoint in- and out-branchings.

Bang-Jensen and Yeo generalized the above conjecture as follows.\footnote{Every strong digraph has an out- and in-branching.}
A digraph $D=(V,A)$ has a {\em good decomposition} if $A$ has two disjoint sets $A_1$ and $A_2$ such that both
$(V,A_1)$ and $(V,A_2)$ are strong \cite{Bang-Jensen-Huang}. 

\begin{conj}\label{conj02}\cite{Bang-Jensen-Yeo}
There exists an integer $N$ so that every $N$-arc-strong digraph $D$
contains a pair of arc-disjoint strong spanning subdigraphs.
\end{conj}

For a general digraph $D$, it is a hard problem to decide whether
$D$ has a decomposition into two strong spanning subdigraphs.

\begin{thm}\label{thm07}\cite{Bang-Jensen-Yeo}
It is NP-complete to decide whether a digraph contains a pair of
arc-disjoint strong spanning subdigraphs.
\end{thm}

Clearly, every digraph with a good decomposition is 2-arc-strong.
Bang-Jensen and Yeo characterized the semicomplete digraphs with a
good decomposition.

\begin{thm}\label{thm03}\cite{Bang-Jensen-Yeo}
A 2-arc-strong semicomplete digraph $D$ has a pair of arc-disjoint
strong spanning subdigraphs if and only if $D$ is not isomorphic to
$S_4$, where $S_4$ is obtained from the complete digraph with four
vertices by deleting a cycle of length four. Furthermore, a good
decomposition of $D$ can be obtained in polynomial time when it
exists.
\end{thm}

The following result extends Theorem \ref{thm03} to locally
semicomplete digraphs.

\begin{thm}\label{thm04}\cite{Bang-Jensen-Huang}
A 2-arc-strong locally semicomplete digraph $D$ has a pair of
arc-disjoint strong spanning subdigraphs if and only if $D$ is not
the second power of an even cycle.
\end{thm}




Let $T$ be a digraph with $t$ vertices $u_1,\dots , u_t$ and let $H_1,\dots H_t$ be digraphs such that $H_i$ has vertices $u_{i,j_i},\ 1\le j_i\le n_i.$
Then the {\em composition} $Q=T[H_1,\dots , H_t]$ is a digraph with vertex set $\{u_{i,j_i}\mid 1\le i\le t, 1\le j_i\le n_i\}$ and arc set
$$A(Q)=\cup^t_{i=1}A(H_i)\cup \{u_{ij_i}u_{pq_p}\mid u_iu_p\in A(T), 1\le j_i\le n_i, 1\le q_p\le n_p\}.$$ 

In this paper, we continue research on good decompositions in
classes of digraphs and consider digraph compositions and products. 

In Section \ref{sec:2}, for digraph compositions $Q=T[H_1,\dots H_t]$, we obtain sufficient conditions for 
$Q$ to have a good decomposition (Theorem \ref{comp-good}) and a characterization
of $Q$ with a good decomposition when $T$ is a strong semicomplete digraph and 
each $H_i$ is an arbitrary digraph with at least two vertices (Theorem \ref{good-decompSemi}). 
Remarkably, in Theorem \ref{good-decompSemi} as in Theorem \ref{thm03},
there are only a finite number of exceptional digraphs, which for Theorem \ref{good-decompSemi} is three. 
Thus, as Theorems \ref{thm03} and \ref{thm04}, Theorem \ref{good-decompSemi} confirms Conjecture \ref{conj02} for a special class of digraphs.

In Section \ref{sec:3}, for digraph products, we prove the following: (a) if $k\geq 2$ is an integer and $G$ is a strong digraph which arcs can be partitioned
into cycles, then the Cartesian product digraph $G^{\square k}$ (the $k$th powers with respect to Cartesian product) has a good decomposition
(Theorem \ref{thm4});
(b) for any strong digraphs $G, H$, the strong product $G\boxtimes H$ has a good decomposition (Theorem \ref{thm2}). 
Necessary definitions of the digraph products are given in Section \ref{sec:3}.

Simple examinations of our constructive proofs show that all our decompositions can be found in polynomial time.

We conclude the paper in Section \ref{sec:4}, where we pose a number of open problems. 

\section{Compositions of digraphs}\label{sec:2}

Let $H'_i$ denote $H_i$ with all arcs deleted, where $1\le i\le t$ and let $Q'=T[H'_1,\dots ,H'_t].$

Compositions of digraphs is a useful concept in digraph theory, see e.g. \cite{Bang-Jensen-Gutin}. In particular, they are used in the Bang-Jensen-Huang characterization
of quasi-transitive digraphs and its structural and algorithmic applications for quasi-transitive digraphs and their extensions, see e.g. \cite{Bang-Jensen-Gutin,Bang-Jensen-Gutin2,GSHC}.

Let us start from a simple observation, which will be useful in the proofs of the theorems of this section.

\begin{lem}\label{lemo}
Let $Q=D[H_1,\dots , H_t].$ If an induced subdigraph $Q^*$ of $Q'=D[H'_1,\dots , H'_t]$ with at least one vertex in each $H_i$
has a good decomposition, then so have $Q'$ and $Q.$
\end{lem}
\begin{pf}
For every $1\le i\le t,$ let $H_i^{(m_i)}$ be the subdigraph of $H'_i$ induced by $\{u_{i,1},u_{i,2},\dots ,u_{i,m_i}\},$ where 
$1\le m_i\le n_i.$
Without loss of generality, let $Q^*=D[H_1^{(m_1)},\dots , H^{(m_t)}_t]$ and let $Q^*$ have a decomposition into arc-disjoint strong spanning subdigraphs $D_1,D_2.$ To extend this decomposition to $Q',$ 
for every $1\le i\le t$ and $j=1,2$, add to $D_j$ the vertices $u_{i,m_i+1},\dots ,u_{i,n_i}$ and let them have the same in- and out-neighbors as $u_{i,1}.$
\end{pf}



The following theorem gives sufficient conditions for a digraph composition to have a good decomposition. As in Theorem \ref{thm03}, $S_4$ will denote the digraph obtained from the complete digraph of order 4 by deleting a cycle of length 4.

\begin{thm}\label{comp-good}
Let $Q=T[H_1,\dots ,H_t],$ where $t\ge 2.$ Then $Q$ has a good decomposition if at least one of the following conditions holds:

(a) $T$ is a 2-arc-strong semicomplete digraph and $H_1,\dots , H_t$ are arbitrary digraphs, but $Q$ is not isomorphic to $S_4;$

(b) $T$ has a Hamiltonian cycle and either $t$ is even and $n_i\ge 2$ for every $i=1,\dots ,t$ or $t$ is odd and $n_i\ge 3$ for every $i=1,\dots ,t$ apart from one $i$ for which $n_i\ge 2,$
or $t$ is odd, $n_i\ge 2$ for every $i=1,\dots ,t$ and at least two distinct subdigraphs $H_i$ have arcs.

(c) If $T$ and all $H_i$ are strong digraphs of orders at least 2.
\end{thm}
\begin{pf}
\noindent{\bf Part (a)}
If $T$ is not isomorphic to $S_4$ then we are done by Theorem \ref{thm03} and Lemma \ref{lemo}.
Now assume that $T$ is isomorphic to $S_4$, but $Q$ is not isomorphic to $S_4$.  Let the vertices of $T$ be $u_1,u_2,u_3,u_4$ and its arcs $$u_1u_2,u_2u_1,u_3u_4,u_4u_3,u_1u_4,u_2u_3,u_4u_2,u_3u_1.$$ Since $Q$ is not isomorphic to $S_4$, at least one of $H_1,H_2,H_3,H_4$ has at least two vertices.
Without loss of generality, let $H_1$ have at least two vertices.
Consider the subdigraph $Q^*$ of $Q'$ induced by $\{u_{1,1}, u_{1,2}, u_{2,1}, u_{3,1}, u_{4,1}\}.$ 
Then $Q^*$ has two arc-disjoint strong spanning subdigraphs: $D_1$ with arcs $$\{u_{1,1}u_{2,1},u_{2,1}u_{1,2},u_{1,2} u_{4,1}, u_{4,1}u_{3,1},u_{3,1}u_{1,1}\}$$ and  $D_2$ with arcs
$$\{u_{2,1}u_{1,1},u_{1,1}u_{4,1},u_{4,1}u_{2,1},u_{2,1}u_{3,1},u_{3,1}u_{1,2},u_{1,2}u_{2,1}\}.$$
It remains to apply Lemma \ref{lemo} to obtain a good decomposition of $Q.$

 \vspace{3mm}
\noindent{\bf Part (b)} Without loss of generality, assume that $u_1u_2\dots u_tu_1$ is a Hamiltonian cycle of $T.$ Let $U=\cup_{i=1}^t\{u_{i,1},u_{i,2}\}.$ 

\paragraph{Case 1: $t$ is even and $n_i\ge 2$ for every $i=1,\dots ,t.$}

The following arc sets induce arc-disjoint strong  spanning subdigraphs $D_1,D_2$ of $Q'[U]:$

\begin{equation}\label{eq1}
\{u_{i,j}u_{i+1,j}\mid 1\le i\le t-1, 1\le j\le 2\}\cup \{u_{t,1}u_{1,2},u_{t,2}u_{1,1}\}
\end{equation}
\begin{equation}\label{eq2}
\{u_{i,j}u_{i+1,(j+1 \mbox{ mod } 2)}\mid 1\le i\le t-1, 1\le j\le 2\}\cup \{u_{t,1}u_{1,1}, u_{t,2}u_{1,2}\}.
\end{equation}

It remains to apply Lemma \ref{lemo}.


 \paragraph{Case 2: $t$ is odd, $n_i\ge 2$ for every $i=1,\dots ,t$ and at least two distinct subdigraphs $H_i$ have arcs.}
Let $e_p,e_q$ be arcs in two distinct subdigraphs $H_p$ and $H_q$. We may assume that both end-vertices of $e_p$ and $e_q$ are in $U.$
Observe that while $D_1$ (with arcs listed in (\ref{eq1})) is strong, $D_2$ (with arcs listed in (\ref{eq2})) forms two arc-disjoint cycles $C$ and $Z.$
We may assume that the tail (head) of $e_p$ ($e_q$) is in $C$ and and the head (tail) of $e_p$ ($e_q$) is in $Z$ (otherwise, relabel vertices in $\{u_{p,1},u_{p,2}\}$ and/or $\{u_{q,1}u_{q_2}\}$).
Thus, adding $e_p$ and $e_q$ to $D_2$ makes it strong.
To obtain two arc-disjoint strong spanning subdigraphs of $Q$ from $D_1,D_2$,
let every vertex $u_{i,j}$ for $j\ge 3$ and $1\le i\le t$ have the same out- and in-neighbors as $u_{i,1}$ in $Q'.$

\paragraph{Case 3: $t$ is odd and $n_i\ge 3$ for every $i=1,\dots ,t$ apart from one $i$ for which $n_i\ge 2.$}
Without loss of generality, assume that $n_1\ge 2$ and $n_i\ge 3$ for all $2\le i\le t.$

First we consider the subcase in which $t=3$, $n_1=2,$ and $n_2=n_3=3.$ Then $Q'$ has two arc-disjoint spanning subdigraphs $D_1$ and $D_2$ with arc sets
$$\{u_{1,1}u_{2,1},u_{3,1}u_{1,1},u_{1,2}u_{2,2},u_{1,2}u_{2,3},u_{3,2}u_{1,2}, u_{3,3}u_{1,2},u_{2,1}u_{3,2},u_{2,2}u_{3,1},u_{2,3}u_{3,3}\},
$$
$$\{u_{1,1}u_{2,2},u_{1,1}u_{2,3},u_{3,2}u_{1,1},u_{3,3}u_{1,1},u_{1,2}u_{2,1},u_{3,1}u_{1,2},u_{2,1}u_{3,3},u_{2,2}u_{3,2},u_{2,3}u_{3,1}\},
$$ respectively. 
It is not hard to see that $D_1$ and $D_2$ are strong by constructing closed walks through all vertices.

Now we extend the previous subcase to that in which $n_1=2$ and
$n_i=3$ for all $2\le i\le t.$ First replace index 3 in every vertex of the form $u_{3,i}$ by $t$ in 
the two arc sets of the previous subcase. 
Then replace every arc of the form
$u_{2,i}u_{t,j}$ in $D_1$ by the path
$u_{2,i}u_{3,i}\dots u_{t-1,i}u_{t,j}.$ In $D_2$, we
replace $u_{2,1}u_{t,3}$ by the path $u_{2,1}u_{3,2}u_{4,1}u_{5,2} \dots
u_{t-1,1}u_{t,3}$, replace $u_{2,2}u_{t,2}$ by the path
$u_{2,2}u_{3,1}u_{4,2}u_{5,1} \dots u_{t-1,2}u_{t,2}$, replace
$u_{2,3}u_{t,1}$ by the path $u_{2,3}u_{3,2}u_{4,3}u_{5,2}\dots
u_{t-1,3}u_{t,1}$, and finally add the path
$u_{2,2}u_{3,3}u_{4,2}u_{5,3}\dots u_{t-1,2}$.

Finally, we extend  the previous subcase to the general one using Lemma \ref{lemo}.

\vspace{3mm}
\noindent{\bf Part (c)} For $j=1,2$, let $T_j$ be the subdigraph of $Q$ induced  by vertex set
$\{u_{i,j}\mid 1\leq i\leq t\}$.
Clearly, $T_1\cong T_2\cong T$ and $T_1$ and $T_2$ are
strong.

Let $Q_1$ be the spanning subdigraph of $Q$ with arc set
$A(Q_1)=A(T_1)\cup(\bigcup_{i=1}^t{A(H_i)})$. Observe that $Q_1$ is
strong since $T_1$ and each $H_i$ are strong, and $T_1$ has a common
vertex with each $H_i$, where $1\leq i\leq t$.

Let $Q_2$ be the spanning subdigraph of $Q$ with arc set
$A(Q_2)=A(Q)\setminus A(Q_1)$. To see that $Q_2$ is strong, we only
need to find a strong subdigraph in $Q_2$ which contains $x$ and $y$
for each pair of distinct vertices $x$ and $y$ in $Q_2$. We will
consider two cases.

\paragraph{Case 1: $x\in V(T_1)$.} Without loss of generality, we assume that $x=u_{1,1}$ and $y\in
\{u_{1,2}, u_{2,1}, u_{2,2}\}$. We first consider the subcase that
$y=u_{2,1}$. Observe that there is at least one arc entering and one arc
leaving $u_{1,2}~(u_{2,2})$ in $T_2$, and so there are two arcs, say
$a$ and $b$~($c$ and $d$), with opposite directions between $x~(y)$
and $T_2$ in $Q_2$. Then by adding the arcs $a, b, c, d$, and the
vertices $x, y$ to $T_2$, we obtain a strong subdigraph $T_2'$ of
$Q_2$ which contains both $x$ and $y$, as desired. For the case that
$y\in \{u_{1,2}, u_{2,2}\}$, we just add the arcs $a, b$, and the
vertex $x$ to $T_2$, and then obtain a strong subdigraph $T_2''$ of
$Q_2$ which contains both $x$ and $y$.

\paragraph{Case 2:  $x\not\in V(T_1)$.} Without loss of generality, we assume that $x=u_{1,2}$ and $y\in
\{u_{1,1}, u_{2,1}, u_{1,3}, u_{2,2}, u_{2,3}\}$ (if $u_{1,3}$ and
$u_{2,3}$ exist). By Case 1 and the fact that $T_2\cong T$ is
strong, we are done if $y\in \{u_{1,1}, u_{2,2}\}$. For the case
that $y= u_{2,1}$, by adding the arcs $c, d$ and the vertex $y$ to
$T_2$, we can obtain a strong subdigraph $T_2'''$ of $Q_2$ which
contains both $x$ and $y$. With a similar argument, we can get the
desired strong subdigraph for the case that $y\in \{u_{1,3},
u_{2,3}\}$.

Hence, we complete the argument and conclude that $Q$ has a good
decomposition.
\end{pf}

We will use Theorem \ref{comp-good} to prove the following characterization for certain compositions $T[H_1,\dots ,H_t]$, where $T$ is a strong semicomplete digraph. 
In the characterization, $\overline{K_p}$ will stand for the digraph of order $p$ with no arcs. Also, $\overrightarrow{C}_k$ and $\overrightarrow{P}_k$ will denote the cycle and path with $k$ vertices, respectively. 

\begin{thm}\label{good-decompSemi}
Let $T$ be a strong semicomplete digraph on $t\ge 2$ vertices and let $H_1,\dots ,H_t$ be arbitrary digraphs, each with at least two vertices.
Then $Q=T[H_1,\dots ,H_t]$ has a good decomposition if and only if $Q$ is not isomorphic to one of the following three digraphs: 
$\overrightarrow{C}_3[\overline{K_2},\overline{K_2},\overline{K_2}]$, $\overrightarrow{C}_3[\overrightarrow{P_2},\overline{K_2},\overline{K_2}].$
$\overrightarrow{C}_3[\overline{K_2},\overline{K_2},\overline{K_3}].$  
\end{thm}
\begin{pf}
Let us first prove the `only if' part of the theorem, i.e. $\overrightarrow{C}_3[\overline{K_2},\overline{K_2},\overline{K_2}],$ 
$\overrightarrow{C}_3[\overrightarrow{P}_2,\overline{K_2},\overline{K_2}]$ and $\overrightarrow{C}_3[\overline{K_2},\overline{K_2},\overline{K_3}]$  do not have good decompositions. By Lemma \ref{lemo}, it suffices to show that neither $\overrightarrow{C}_3[\overrightarrow{P}_2,\overline{K_2},\overline{K_2}]$ nor $\overrightarrow{C}_3[\overline{K_2},\overline{K_2},\overline{K_3}]$ has a good decomposition.
The proof is by reductio ad absurdum. 


Suppose that $Q=\overrightarrow{C}_3[\overrightarrow{P_2},\overline{K_2},\overline{K_2}]$ has a decomposition into two strong spanning subdigraphs $Q_1,Q_2.$
Since $Q$ has 13 arcs, without loss of generality, we may assume that $Q_1$ is a Hamiltonian cycle of $Q.$ Since the arc of $H_1$ cannot be in a Hamiltonian cycle of $Q$, without loss of generality, let $Q_1=u_{1,1}u_{2,1}u_{3,1}u_{1,2}u_{2,2}u_{3,2}u_{1,1}$. Then the remaining arcs of $Q$ form two disjoint cycles $u_{1,1}u_{2,2}u_{3,1}u_{1,1}$ and $u_{1,2}u_{2,1}u_{3,2}u_{1,2}$ and a single arc between them, a contradiction to the assumption that $Q_2$ is strong. 

Suppose that $Q=\overrightarrow{C}_3[\overline{K_2},\overline{K_2},\overline{K_3}]$  has a decomposition into two strong spanning subdigraphs $Q_1,Q_2.$ Since $Q$ has 16 arcs and has no Hamiltonian cycle, each of  $Q_1,Q_2$ has 8 arcs. Since $Q$ has only cycles of lengths 3 and 6 and $Q_1$ is strong, without loss of generality, we may assume that $Q_1$ consists of a cycle $u_{1,1}u_{2,1}u_{3,1}u_{1,2}u_{2,2}u_{3,2}u_{1,1}$ and a path $u_{2,1}u_{3,3}u_{1,1}.$ Then $Q_2$ consists of two cycles $u_{1,1}u_{2,2}u_{3,1}u_{1,1}$ and $u_{1,2}u_{2,1}u_{3,2}u_{1,2}$ and a path $u_{2,2}u_{3,3}u_{1,2}.$ 
Observe that $Q_2$ is not strong, a contradiction.

\vspace{3mm}

Now we will show the `if' part of the theorem by reductio ad absurdum as well.  
Assume that $Q$ is not isomorphic to either of the three digraphs, but has no good decomposition. 

By Camion's Theorem \cite{Camion}, $T$ has a Hamiltonian cycle $C=u_1u_2\dots u_tu_1$. Thus, Conditions (b) of Theorem \ref{comp-good} are applicable.
By the conditions, $t$ must be odd and for at least two distinct indexes $p,q\in \{1,2,\dots ,t\}$, we have $n_p=n_q=2.$

Suppose $t\ge 5.$ Then there will be arcs between $H_i$ and $H_{i+2}$ in $Q$ for every $i=1,2,\dots ,t-2.$ Recall Case 2 of Part (b) of the proof of Theorem \ref{comp-good}. 
The arcs between $H_i$ and $H_{i+2}$ arcs can be used to make $D_2$ strong instead of arcs  $e_p$ and $e_q$ used in 
Case 2 of Part (b) of the proof of Theorem \ref{comp-good}.
Thus, $Q$ has a good decomposition, a contradiction.  Hence, $t=3$ and, without loss of generality, $n_1=n_2=2$ and $n_3\ge 2.$ 

Suppose that $T$ has opposite arcs. One of these arcs will not be on the Hamiltonian cycle  $C$ of $T$ and will correspond to four or more arcs in $Q.$ Now
recall Case 2 of Part (b) of the proof of Theorem \ref{comp-good}. Two of the above-mentioned arcs can be used to make $D_2$ strong instead of arcs  
$e_p$ and $e_q$ used in Case 2 of Part (b) of the proof of Theorem \ref{comp-good}. Thus, $Q$ has a good decomposition, a contradiction.  Hence, 
$T=\overrightarrow{C}_3.$ 

Suppose that $n_3\ge 4.$ To get a contradiction, by Lemma \ref{lemo} it suffices to show that $Q=\overrightarrow{C}_3[\overline{K_2},\overline{K_2},\overline{K_4}]$ has a decomposition into two strong spanning subdigraphs $D_1,D_2,$ where 
$D_1$ consists of a cycle $u_{1,1}u_{2,1}u_{3,1}u_{1,2}u_{2,2}u_{3,2}u_{1,1}$ and two paths $u_{2,1}u_{3,4}u_{1,1}$ and $u_{2,2}u_{3,3}u_{1,2}$ and
$D_2$ consists of   two cycles $u_{1,1}u_{2,2}u_{3,1}u_{1,1}$ and $u_{1,2}u_{2,1}u_{3,2}u_{1,2}$ and two paths $u_{2,1}u_{3,3}u_{1,1}$ and $u_{2,2}u_{3,4}u_{1,2}.$
Thus, $n_3\le 3.$

Now consider the case of $n_1=n_2=2$ and $n_3=3.$ Since $Q$ is not isomorphic to $\overrightarrow{C}_3[\overline{K_2},\overline{K_2},\overline{K_3}],$ it has an arc in either $H_1$ or $H_2$ or $H_3$, and by Conditions (b) of Theorem \ref{comp-good}, only one of $H_1, H_2, H_3$ has an arc $a.$ 
Without loss of generality, assume that if $H_1$ has an arc then $a=u_{1,2}u_{1,1}$, if $H_2$ has an arc then $a=u_{2,1}u_{2,2}$ and if $H_3$ has an arc then $a=u_{3,2}u_{3,1}.$
Then $Q$ has a decomposition into two spanning subdigraphs $D_1,D_2,$ where $D_1$ consists of a cycle $u_{1,1}u_{2,1}u_{3,1}u_{1,2}u_{2,2}u_{3,2}u_{1,1}$ and a path $u_{2,1}u_{3,3}u_{1,1}$ and $D_2$ consists of two cycles $u_{1,1}u_{2,2}u_{3,1}u_{1,1}$ and $u_{1,2}u_{2,1}u_{3,2}u_{1,2}$, a path $u_{2,2}u_{3,3}u_{1,2}$ and arc $a$. Observe that both $D_1$ and $D_2$ are strong, a contradiction.

It remains to consider the case of $n_1=n_2=n_3=2.$  Since $Q$ is not isomorphic to $\overrightarrow{C}_3[\overline{K_2},\overline{K_2},\overline{K_2}]$, at least one of $H_1,H_2$ and $H_3$ has an arc. By Conditions (b) of Theorem \ref{comp-good}, only one of $H_1,H_2$ and $H_3$ has an arc. Without loss of generality, assume that $H_1$ has an arc.
Suppose that $H_1$  has two arcs. Then $H_1=\overrightarrow{C}_2$. Then we can use the arcs of $H_1$ to make $D_2$ strong instead of arcs  
$e_p$ and $e_q$ used in Case 2 of Part (b) of the proof of Theorem \ref{comp-good}. Thus, $Q$ has a good decomposition, a contradiction.  
Hence,  if $H_1$ has an arc, it must have just one arc. This concludes our proof. 
\end{pf}

\vspace{2mm}


\section{Products of digraphs}\label{sec:3}

The {\em Cartesian product} $G\square H$ of two digraphs $G$ and $H$
is a digraph with vertex set $V(G\square H)=V(G)\times V(H)=\{(x,
x')\mid x\in V(G), x'\in V(H)\}$ and arc set $A(G\square
H)=\{(x,x')(y,y')\mid xy\in A(G), x'=y',~or~x=y, x'y'\in A(H)\}.$ By
definition, we know the Cartesian product is associative and
commutative, and $G\square H$ is strongly connected if and only if
both $G$ and $H$ are strongly connected \cite{Hammack}. We define
the $n$th powers with respect to Cartesian product as $D^{\square
n}=D\square D\square \cdots \square D$.

\begin{figure}[!hbpt]
\begin{center}
\includegraphics[scale=0.8]{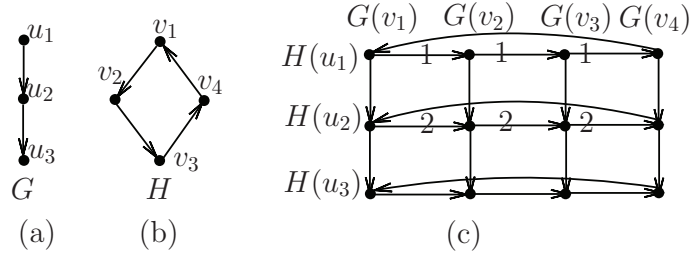}
\end{center}
\caption{Two digraphs $G$, $H$ and their Cartesian
product.}\label{figure1}
\end{figure}

In the argument of this section, we will use the following
terminology and notation. Let $G$ and $H$ be two digraphs with
$V(G)=\{u_i \mid 1\leq i\leq n\}$ and $V(H)=\{v_j \mid 1\leq j\leq
m\}$. For simplicity, we let $u_{i,j}=(u_i,v_j)$ for $1\leq i\leq n,
1\leq j\leq m$. We use $G(v_j)$ to denote the subdigraph of
$G\square H$ induced by vertex set $\{u_{i,j}\mid 1\leq i\leq n\}$
where $1\leq j\leq m$, and use $H(u_i)$ to denote the subdigraph of
$G\square H$ induced by vertex set $\{u_{i,j}\mid 1\leq j\leq m\}$
where $1\leq i\leq n$. Clearly, we have $G(v_j)\cong G$ and
$H(u_i)\cong H$. (For example, as shown in Figure \ref{figure1},
$G(v_j)\cong G$ for $1\leq j\leq 4$ and $H(u_i)\cong H$ for $1\leq
i\leq 3$.) For $1\leq j_1\neq j_2\leq m$, $u_{i,j_1}$ and
$u_{i,j_2}$ belong to the same digraph $H(u_i)$ where $u_i\in V(G)$;
we call $u_{i,j_2}$ the {\em vertex corresponding to} $u_{i,j_1}$ in
$G(v_{j_2})$; for $1\leq i_1\neq i_2\leq n$, we call $u_{i_2,j}$
the vertex corresponding to $u_{i_1,j}$ in $H(u_{i_2})$. Similarly,
we can define the subdigraph {\em corresponding} to some other
subdigraph. For example, in Fig. \ref{figure1}(c), let $P_1$~$(P_2)$
be the path labelled 1 (2) in $H(u_1)~(H(u_2))$, then $P_2$ is
called the path {\em corresponding} to $P_1$ in $H(u_2)$.

\begin{lem}\label{lem1}
For any integer $n\geq 2$, the product digraph
$D=\overrightarrow{C}_n\square \overrightarrow{C}_n$ can be
decomposed into two arc-disjoint Hamiltonian cycles.
\end{lem}
\begin{pf} Let $G=H\cong \overrightarrow{C}_n$;
moreover $G=u_1u_2\dots u_nu_1$ and $H=v_1v_2\dots v_nv_1.$
Let $P_i=G(v_i)-u_{n-i,i}u_{n+1-i,i}$ for
$1\leq i\leq n-1$ and $P_n=G(v_n)-u_{n,n}u_{1,n}$.
Let $D'$ be the subdigraph of $D$ which is a union of $n$ paths $P_i$
and the following $n$ arcs:
$\{u_{n-i,i}u_{n-i,i+1}\mid 1\leq i\leq n-1\}\cup
\{u_{n,n}u_{n,1}\}.$  Let $D''$ be a
spanning subdigraph of $D$ with $A(D'')=A(D)\setminus A(D')$. It is
not hard to check that both $D'$ and $D''$ are Hamiltonian cycles of $D$;
this completes the proof.
\end{pf}

Note that deciding whether a digraph $D$ has a collection of arc-disjoint cycle covering all vertices of $D$, can be done in polynomial time using network flows.
Indeed, assign lower bound 1 and upper bound $\min\{d^-(x),d^+(x)\}$ to every vertex $x$ in $D$ and lower bound 0 and upper bound 1 to every arc of $D$.
Observe that the resulting network has a feasible flow if and only if $D$ has a collection of arc-disjoint cycle covering all vertices of $D$. Observe that the existence of
a flow in a network with lower and upper bounds on vertices and arcs can be decided in polynomial time, see e.g. Chapter 4 in \cite{Bang-Jensen-Gutin}. Moreover, we can compute such a flow in polynomial time (if it exists) and obtain the corresponding collection of cycles in $D.$
The following lemma may be of independent interest.

\begin{lem}\label{lem4}
Let $G$ be a strong digraph of order at least two which has a collection of arc-disjoint cycle covering all its vertices.
Then the
product digraph $D=G\square G$ can be decomposed into two
arc-disjoint strong spanning subdigraphs. Moreover, these two
arc-disjoint strong spanning subdigraphs can be found in polynomial
time.
\end{lem}
\begin{pf}
By the arguments in the paragraph before this lemma, we may assume
that we are given a collection $(P_0, P_1, P_2, \cdots, P_p)$ of
arc-disjoint cycle covering all vertices of $G$. For each
$h=0,1,2,\cdots,p$, let $G_h$ denote the digraph with vertices
$\bigcup_{i=0}^h{V(P_i)}$ and arcs $\bigcup_{i=0}^h{A(P_i)}$. Now we
will prove the lemma by induction on the number of cycles in the
collection.

For the base step, by Lemma \ref{lem1}, we have that $G_0\square
G_0=P_0\square P_0$ can be decomposed into two arc-disjoint strong spanning
subdigraphs.

For the inductive step, we assume that $G_h\square G_h~(0\leq h\leq
p-1)$ can be decomposed into two arc-disjoint strong spanning
subdigraphs $D_h'$ and $D_h''$. We will construct two arc-disjoint
strong spanning subdigraphs in $G_{h+1}\square G_{h+1}$.

If $V(G_h)\subseteq V(P_{h+1})$, then $P_{h+1}$ is a Hamiltonian
cycle of $G_{h+1}$, and we are done by Lemma \ref{lem1}. If
$V(P_{h+1})\subseteq V(G_h)$, then $G_h$ is a strong spanning
subdigraph of $G_{h+1}$, and we are also done.

In the following argument, we assume that $V(G_h)\setminus
V(P_{h+1})\neq \emptyset$ and $V(P_{h+1})\setminus V(G_h)\neq
\emptyset$. Without loss of generality, for the first copies of
$G_h$ and $P_{h+1}$ in $G_h\square G_h$ and $P_{h+1}\square
P_{h+1}$, let $V(G_h)=\{u_i\mid 1\leq i\leq t\}$,
$V(P_{h+1})=\{u_i\mid s\leq i\leq \ell\}$. We have $1<s\leq t<\ell$.
For the second copies of $G_h$ and $P_{h+1}$ in $G_h\square G_h$ and
$P_{h+1}\square P_{h+1}$, we will use $v_i$'s rather than $u_i$'s.

By Lemma \ref{lem1}, in $G_{h+1}\square G_{h+1}$, the subdigraph
$P_{h+1}\square P_{h+1}$ can be decomposed into two arc-disjoint
strong spanning subdigraphs $\overline{D}_h'$ and
$\overline{D}_h''$. Observe that $$V(G_h\square G_h)\cap
V(P_{h+1}\square P_{h+1})\supseteq \{u_{t,t}\} \mbox{ and  } A(G_h\square G_h)\cap A(P_{h+1}\square
P_{h+1})=\emptyset.$$ For $1\leq j\leq s-1$, let $G_{h,j}$ be the
subdigraph of $G(v_j)$ corresponding to $P_{h+1}$. For $t+1\leq
j\leq \ell$, let $G_{h,j}$ be the subdigraph of $G(v_j)$
corresponding to $G_h$. For $1\leq i\leq s-1$, let $H_{h,i}$ be the
subdigraph of $H(u_i)$ corresponding to $P_{h+1}$. For $t+1\leq
i\leq \ell$, let $H_{h,i}$ be the subdigraph of $H(u_i)$
corresponding to $G_h$.

Now let $D_{h+1}'$ be a union of the following strong digraphs:
$D_h'$, $\overline{D}_h'$, $H_{h,i}$ and $G_{h,j}$ for all $t+1\leq
i,j\leq \ell$. Observe that $D_{h+1}'$ is a strong spanning
subdigraph of $G_{h+1}\square G_{h+1}$ since $\overline{D}_h'$ has
at least one common vertex with each of $D_h'$, $H_{h,i}$ and
$G_{h,j}$ for all $t+1\leq i,j\leq \ell$.
 Let $D_{h+1}''$ be a spanning subdigraph
of $G_{h+1}\square G_{h+1}$ with $A(D_{h+1}'')=A(G_{h+1}\square
G_{h+1})\setminus A(D_{h+1}')$. Observe that $D_{h+1}''$ is the
union of $D_h''$, $\overline{D}_h''$, $H_{h,i}$ and $G_{h,j}$ for
all $1\leq i,j\leq s-1$. And $D_h''$ has at least one common vertex
with each of $\overline{D}_h''$, $H_{h,i}$ and $G_{h,j}$ for all
$1\leq i,j\leq s-1$, thus $D_{h+1}''$ is strong.

Hence, we complete the inductive step and conclude that $D=G\square
G$ can be decomposed into two arc-disjoint strong spanning
subdigraphs. Moreover, by the above argument, these subdigraphs can
be found in polynomial time.
\end{pf}

\begin{lem}\label{lem5}
For any two strong digraphs $G$ and $H$, if $G$ contains a pair of
arc-disjoint strong spanning subdigraphs, then the product digraph
$D=G\square H$ can be decomposed into two arc-disjoint strong
spanning subdigraphs.
\end{lem}
\begin{pf}
Let $V(G)=\{u_i \mid 1\leq i\leq n\}, V(H)=\{v_j \mid 1\leq j\leq
m\}$, and $G$ contain two arc-disjoint strong spanning subdigraphs
$G_1$ and $G_2$. For $1\leq j\leq m$, let $G_{1,j}$ be the
subdigraph of $G(v_j)$ corresponding to $G_1$. Let $D'$ be a union
of $H(u_1)$ and $G_{1,j}$ for all $1\leq j\leq m$, and $D''$ be a
subdigraph of $D$ with $V(D'')=V(D)$ and $A(D'')=A(D)\setminus
A(D')$. It is not hard to verify that both $D'$ and $D''$ are strong spanning
subdigraphs of $D$. This completes the proof.
\end{pf}

By the definition of $D^{\square k}$, associativity of the
Cartesian product, and Lemmas \ref{lem4} and \ref{lem5}, we can
obtain the following result on $G^{\square k}$ for any integer
$k\geq 2$.
\begin{thm}\label{thm4}Let $G$ be a strong digraph of order at least two which has a collection of arc-disjoint cycle covering all its
vertices and let $k\ge 2$ be an integer. Then the product digraph $D=G^{\square k}$ can be
decomposed into two arc-disjoint strong spanning subdigraphs.
Moreover, for any fixed integer $k$, these two subdigraphs can be
found in polynomial time.
\end{thm}


The {\em strong product} $G\boxtimes H$ of two digraphs $G$ and $H$
is a digraph with vertex set $V(G\boxtimes H)=V(G)\times V(H)=\{(x,
x')\mid x\in V(G), x'\in V(H)\}$ and arc set $A(G\boxtimes
H)=\{(x,x')(y,y')\mid xy\in A(G), x'=y',~or~x=y, x'y'\in
A(H),~or~xy\in A(G),~x'y'\in A(H)\}.$ By definition, $G\square H$ is
a spanning subdigraph of $G\boxtimes H$, and $G\boxtimes H$ is
strongly connected if and only if both $G$ and $H$ are strongly
connected \cite{Hammack}. In the following argument, we will still
use the terminology and notation introduced earlier in this section, since $G\square H$ is
a spanning subdigraph of $G\boxtimes H$.

\begin{lem}\label{lem2}
For any two integers $n,m \geq 2$, the product digraph
$D=\overrightarrow{C}_n\boxtimes \overrightarrow{C}_m$ can be
decomposed into two arc-disjoint strong spanning subdigraphs.
\end{lem}
\begin{pf} Let $\overrightarrow{C}_n=u_1u_2\dots u_nu_1$ and
$\overrightarrow{C}_m=v_1v_2\dots v_m.$ Let $D'$ be the spanning
subdigraph of $D$ which is the union of $G(v_j)$ for $1\leq j\leq m$
and the following additional $m$ arcs: $\{u_{n, j}u_{1,j+1}\mid
1\leq j\leq m-1\}\cup \{u_{1,m}u_{2,1}\}$. Observe that $D'$ is
strong. Let $D''$ be a spanning subdigraph of $D$ with
$A(D'')=A(D)\setminus A(D')$. To see that $D''$ is strong, observe that
it contains $H(u_i)$ for $1\leq i\leq n$ and arcs $\{u_{i,1}
u_{i+1,2}\mid 1\le i\le n-1\}\cup \{u_{n,m}u_{1,1}\}.$
\end{pf}

We will use the following decomposition of strong digraphs.

An {\em ear decomposition} of a digraph $D$ is a sequence
$\mathcal{P}=(P_0, P_1, P_2, \cdots, P_t)$, where $P_0$ is a cycle
or a vertex and each $P_i$ is a path, or a cycle with the following
properties:\\
$(a)$~$P_i$ and $P_j$ are arc-disjoint when $i\neq j$.\\
$(b)$~For each $i=0,1,2,\cdots,t$: let $D_i$ denote the digraph with
vertices $\bigcup_{j=0}^i{V(P_j)}$ and arcs
$\bigcup_{j=0}^i{A(P_j)}$. If $P_i$ is a cycle, then it has
precisely one vertex in common with $V(D_{i-1})$. Otherwise the end
vertices of $P_i$ are distinct vertices of $V(D_{i-1})$ and no other
vertex of $P_i$ belongs to $V(D_{i-1})$.\\
$(c)$~$\bigcup_{j=0}^t{A(P_j)}=A(D)$.

The following result is well-known, see e.g.
\cite{Bang-Jensen-Gutin}.
\begin{thm}\label{thm01}
Let $D$ be a digraph with at least two vertices. Then $D$ is strong
if and only if it has an ear decomposition. Furthermore, if $D$ is
strong, every cycle can be used as a starting cycle $P_0$ for an ear
decomposition of $D$, and there is a linear-time algorithm to find such
an ear decomposition.
\end{thm}

\begin{thm}\label{thm2}
For any strong digraphs $G$ and $H$ with orders at least
2, the product digraph $D=G\boxtimes H$ can be decomposed into two
arc-disjoint strong spanning subdigraphs. Moreover, these two
arc-disjoint strong spanning subdigraphs can be found in polynomial
time.
\end{thm}
\begin{pf}
By Theorem \ref{thm2} $G$ has an ear decomposition $\mathcal{P}=(P_0, P_1, P_2,
\cdots, P_p)$ and $H$ has an ear decomposition $\mathcal{Q}=(Q_0,
Q_1, Q_2, \cdots, Q_q)$, such that $P_0$ is a cycle of $G$ and
$Q_0$ is a cycle of $H$ by Theorem \ref{thm01}. Let $G_i$ denote the
subdigraph of $G$ with vertices $\bigcup_{j=0}^i{V(P_j)}$ and arcs
$\bigcup_{j=0}^i{A(P_j)}$ and let $H_i$ denote the
subdigraph of $H$ with vertices $\bigcup_{j=0}^i{V(Q_j)}$ and arcs
$\bigcup_{j=0}^i{A(Q_j)}.$

We will prove the theorem by induction on $r\in \{0,1,\dots ,p+q\}.$
For the base step, by Lemma \ref{lem2}, we have that $P_0\boxtimes
Q_0$ can be decomposed into two arc-disjoint strong
spanning subdigraphs.
For the inductive step, we assume that $r=h+g<p+q\ (h\le p, g\le q)$ and
$G_h\boxtimes H_g$ can be decomposed into two arc-disjoint strong spanning
subdigraphs $D'$ and $D''$.

Since strong product is a commutative operation, without loss of generality it suffices to prove that
$G_{h+1}\boxtimes H_g$ $(h<p)$ can be decomposed into two arc-disjoint strong spanning
subdigraphs. Let $V(G_h)=\{u_1,u_2,\dots ,u_{\ell}\}$, $V(H_g)=\{v_1,v_2,\dots ,v_m\}$ and $v_1v_s\in A(H_g).$
Let $P_{h+1,j}$ be the subdigraph of $G(v_j)$ corresponding to $P_{h+1}$ for $1\le j\le m.$
We will consider two cases.

\paragraph{Case 1: $P_{h+1}$ is a cycle.} Let $P_{h+1}=u_{\ell}u_{\ell +1}\dots u_nu_{\ell}.$
 Observe that every
$P_{h+1,j}$ for $1\le j\le m$ shares vertex $u_{\ell,j}$ with $D'.$ Thus, the union $U_1$ of $D'$ and $P_{h+1,j}$ for $1\le j\le m$
is a strong spanning subdigraph of $G_{h+1}\boxtimes H_g.$ Let $V(U_2)=V(G_{h+1}\boxtimes H_g)$ and $A(U_2)=A(G_{h+1}\boxtimes H_g)\setminus A(U_1).$

Observe that $A(U_2)$ contains $A(D'')$, $A(H(u_i))$ for $\ell +1\le i\le n$ and $\{u_{i,1}u_{i+1,s}\mid \ell\le i\le n-1\}\cup \{u_{n,1}u_{\ell,s}\}.$ Thus, $U_2$ is strong.

\paragraph{Case 2: $P_{h+1}$ is a path.} Let $P_{h+1}=u_{\ell}u_{\ell +1}\dots u_{n-1}u_t,$ where $t<\ell.$ Let $U_1$ be the union of $D'$ and $P_{h+1,j}$ for $1\le j\le m.$
Observe that $U_1$ is a spanning subdigraph of $G_{h+1}\boxtimes H_g$ and strong since every $P_{h+1,j}$ for $1\le j\le m$ shares its end-vertices with $D'.$
Let $V(U_2)=V(G_{h+1}\boxtimes H_g)$ and $A(U_2)=A(G_{h+1}\boxtimes H_g)\setminus A(U_1).$
Observe that $A(U_2)$ contains $A(D'')$, $A(H(u_i))$ for $\ell +1\le i\le n-1$ and $\{u_{i,1}u_{i+1,s}\mid \ell\le i\le n-2\}\cup \{u_{n-1,1}u_{t,s}\}.$ Thus, $U_2$ is strong.

\vspace{1mm}

Hence, we complete the inductive step and conclude that
$D=G\boxtimes H$ can be decomposed into two arc-disjoint strong
spanning subdigraphs. Furthermore, by Theorem \ref{thm01}, the
proof of Lemma \ref{lem2}, and the argument of this
theorem, we can conclude that these two strong spanning subdigraphs
can be found in polynomial time.
\end{pf}

The {\em lexicographic product} $G\circ H$ of two digraphs $G$ and
$H$ is a digraph with vertex set $V(G\circ H)=V(G)\times V(H)=\{(x,
x')\mid x\in V(G), x'\in V(H)\}$ and arc set $A(G\circ
H)=\{(x,x')(y,y')\mid xy\in A(G),~or~x=y~and~x'y'\in
A(H)\}$\cite{Hammack}. By definition, $G\boxtimes H$ is a
spanning subdigraph of $G\circ H$, so the following result holds by
Theorem \ref{thm2}: For any strong connected digraphs $G$ and $H$
with orders at least 2, the product digraph $D=G\circ H$ can be
decomposed into two arc-disjoint strong spanning subdigraphs.
Moreover, these two arc-disjoint strong spanning subdigraphs can be
found in polynomial time. In fact, we can get a more general result.

A digraph is {\em Hamiltonian decomposable} if it has a family of
Hamiltonian dicycles such that every arc of the digraph belongs to
exactly one of the dicycles. Ng \cite{Ng} gives the most complete
result among digraph products.

\begin{thm}\label{thm06}\cite{Ng}
If $G$ and $H$ are Hamiltonian decomposable digraphs, and $|V(G)|$
is odd, then $G\circ H$ is Hamiltonian decomposable.
\end{thm}

Theorem \ref{thm06} implies that if $G$ and $H$ are Hamiltonian
decomposable digraphs, and $|V(G)|$ is odd, then $G\circ H$ can be
decomposed into two arc-disjoint strong spanning subdigraphs. It is
not hard to extend this result as follows: for any strong digraphs
$G$ and $H$ of orders at least 2, if $H$ contains $\ell \geq 1$
arc-disjoint strong spanning subdigraphs, then the product digraph
$D=G\circ H$ can be decomposed into $\ell +1$ arc-disjoint strong
spanning subdigraphs.


\section{Open Problems}\label{sec:4}

We have characterized digraphs $T[H_1,\dots ,H_t],$ where  $T$ is strong semicomplete and every $H_i$ is arbitrary with at least two vertices, 
which have a good decomposition. It is a natural open problem to extend the characterization to all such digraphs, where some 
$H_i$'s can have just one vertex.  Of course, the extended characterization would generalize also Theorem \ref{thm03}.

A digraph $Q$ is {\em quasi-transitive}, if for any triple $x,y,z$ of distinct vertices of $Q$, if $xy$ and $yz$ are arcs of $Q$ 
then either $xz$ or $zx$ or both are arcs of $Q.$ For a recent survey on quasi-transitive digraphs and their generalizations, see a chapter 
\cite{GSHC} by Galeana-S\'anchez and Hern\'andez-Cruz. 
Bang-Jensen and Huang \cite{Bang-Jensen-Huang95} proved that a quasi-transitive digraph is strong if and only if
$Q=T[H_1,\dots ,H_t],$ where $T$ is a strong semicomplete digraph and each $H_i$ is 
a non-strong quasi-transitive digraph or has just one vertex. 
Thus, a special case of the above problem is to characterize strong quasi-transitive digraphs
with a good decomposition. This would generalize Theorem \ref{thm03} as well.

We believe that these characterizations will confirm Conjecture \ref{conj02} for the classes of quasi-transitive digraphs and digraphs  
$T[H_1,\dots ,H_t],$ where  $T$ is strong semicomplete. In the absence of the characterizations, it would still be interesting to confirm the conjecture
at least for quasi-transitive digraphs.

In Lemma \ref{lem4}, we show that $G\square H$ contains a pair of
arc-disjoint strong spanning subdigraphs when $G\cong H$. However,
the following result implies Lemma
\ref{lem4} cannot be extended to the case that $G\not\cong H$, since it
is not hard to show that the Cartesian product digraph of any two
cycles has a pair of arc-disjoint strong spanning subdigraphs if and
only if it has a pair of arc-disjoint Hamiltonian cycles.

\begin{thm}\label{thm05}\cite{Trotter-Erdos}
The Cartesian product $\overrightarrow{C_p}\square
\overrightarrow{C_q}$ is Hamiltonian if and only if there are
non-negative integers $d_1, d_2$ for which $d_1+d_2= \gcd(p, q)\geq
2$ and $\gcd(p, d_1) = \gcd(q, d_2) = 1$.
\end{thm}

However, Lemma \ref{lem4} could hold for the case that
$G\not\cong H$ if we add other conditions. As shown in Lemma
\ref{lem5}, we know $G\square H$ contains a pair of arc-disjoint
strong spanning subdigraphs when one of $G$ and $H$ contains a pair of
arc-disjoint strong spanning subdigraphs. So the following open question is interesting: for any two strong digraphs $G$ and
$H$, neither of which contain a pair of arc-disjoint
strong spanning subdigraphs, under what condition the product
digraph $G\square H$ contains a pair of arc-disjoint strong spanning
subdigraphs?

Furthermore, we may also consider the following more challenging question: under what conditions the product digraph
$G\square H~(G\boxtimes H)$ has more (than two) arc-disjoint strong spanning
subdigraphs?



\end{document}